\documentclass[onecolumn]{article}

\usepackage{arxiv}
\usepackage[numbers]{natbib}
\usepackage{subcaption}
\usepackage{graphicx}
\usepackage[hidelinks]{hyperref}
\usepackage{url}
\usepackage{placeins}
\usepackage{multirow}
\usepackage{amsmath}
\usepackage{lipsum}

\usepackage{amssymb}

\usepackage{float}

\newcommand\blfootnote[1]{%
	\begingroup
	\renewcommand\thefootnote{}\footnote{#1}%
	\addtocounter{footnote}{-1}%
	\endgroup
}

\begin{document}
\title{THE CHARACTERISTICS AND PROPERTIES OF GROWING NETWORK BASED ON HEXAHEDRON}

\author{
	Li Haijun\footnote{}\\
	Information Technology Centre\\
    Hexi University\\
	Zhangye 734000, P. R. China\\
	\texttt{lihj@hxu.edu.cn} \\
	\And
    Lu Qingping\\
	College of Foreign Languages and Literature\\
	Hexi University\\
    Zhangye 734000, P. R. China\\
	\texttt{luqp@hxu.edu.cn}\\
}

% Uncomment to remove the date
%
% Uncomment to override  the `A preprint' in the header
%\renewcommand{\headeright}{Technical Report}
%\renewcommand{\undertitle}{Technical Report}

\maketitle
%\tableofcontents
\begin{abstract}
We constructs a new network by superposition of hexahedron , which are scale-free, highly sparse, disassortative ,and maximal planar graphs. The network degree distribution, agglomeration coefficient and degree of correlation are computed separately using the iterative method, and these characteristics are found to be very rich. The method of network characteristic analysis can be applied to some actual systems, so as to study the complexity of real network system under the framework of complex network theory.

\end{abstract}

% keywords can be removed
%\smallskip
\noindent \textbf{Keywords.} Study;Characteristics and Properties;Growing Network;Hexahedron. \blfootnote{$^*$ Corresponding author.}

\section{Introduction}
Scientists have studied all aspects of the graph, which is a very interesting job.The graph of discrete objects has been studied extensively for more than 100 years.The analysis of the classification and nature of the graph and the development of graph algorithm are important problems in this direction.In the field of application, discrete graphics can represent physical, biological or sociological objects, such as crystal or protein structure networks, community network, etc.

The way to show that the real world's complex systems use networks associated with vertices and edges has been developed.[1-3]. In particular, some of the characteristics of complex network properties such as high sparseness, scale-free distribution has shown general attributes that plays an important role in determining related special properties of the complex systems and complex phenomena.Inspired by the two pioneering models[4,5],a large number of models have been proposed in the study of network topology, and these models have made great progress[6-11].Thanks to both the advances in network theory and the increasing amount of network data available,the researchers began to reveal a large number of different progress that could lead to some of the aforementioned general characteristics.The research of universal structural properties in the field of complex network modeling is very interesting[1-4].

This paper studies related properties based on the superimposed hexahedron of the network, we built a new network using iterative superimposed methods.The network generated after iteration $n$ times is random or determined,which are maximal planar graphs, show scale-free degree distribution and assortative degree-degree correlations.
\section{Network construction}
We will explore using hexahedron to generate a new network.We extended the idea in Ref.[12] to deal with this problem. This literature introduces the maximal planar graphs and generates a kind of planar network through regular convex hexahedron.Therefore, the framework of bubble topology is readily applicable to the hexahedron. This is illustrated in Fig.1.
\begin{figure*}
	\centering
	\includegraphics[height=2cm,width=6.9cm]{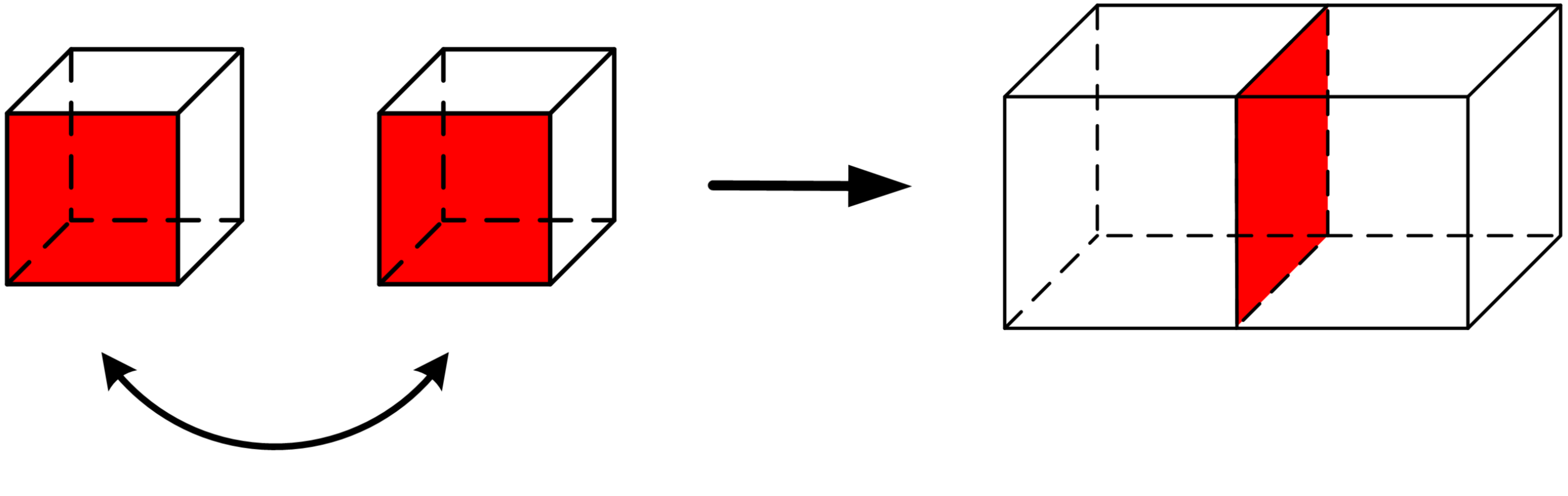}
	\caption{(Color online) This Figure describes how to join two hexahedron by merging the square faces}
	\label{num_histo_pic}
\end{figure*}
We generate a planar network in a determinate way, which iteratively merge the newly introduced bubbles into the square faces of the graph $G(t)$.The plane network generated by deterministic processing is called ``Deterministic Bubble Networks'' (DBNs).These processes are illustrated in Fig.2.
\begin{figure*}
	\centering
	\includegraphics[height=2cm,width=6.9cm]{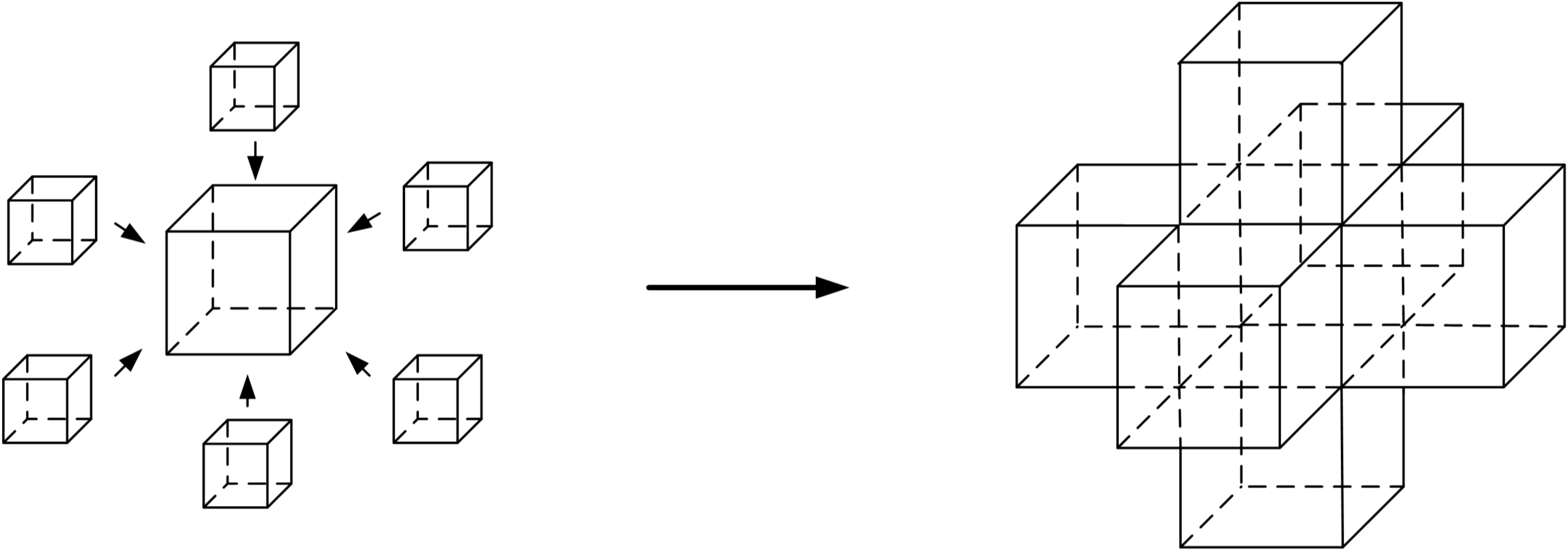}
	\caption{Process of merging newly introduced bubbles to all polygonal faces of a growing graph}
	\label{num_histo_pic}
\end{figure*}
We note that these processes can be seen as a generalization of the Apollonian network construction process[13,14] and the Sierpinski gasket network of trisecting sides [14].The essence of the Apollo network is the deterministic network,it is generated by tetrahedral bubbles.The difference that distinguishes the hexahedron DBN from Apollonian network is that the Apollonian network considers only three triangular faces of the initial tetrahedron,whereas the bubble approach considers all six faces of initial hexahedron.
\section{Iterative algorithm of the network}
In the construction process of the hexahedron network,each face of the original hexahedron is joined by a quadrilateral at the first time, and the vertices of the new quadrilateral are connected to the vertices of the original quadrilateral.The second generation of the new quadrangle vertex is connected to the last generation of quadrilateral vertices, so that the network is generated over and over again.When building a network, each group has four new nodes that are added, and five new quadrilaterals are generated, each of which can create four nodes in the next generation.In this idea we can introduce an iterative algorithm to create the corresponding network,denoted by $H(t)$ after $t$ generation evolutions.

The iterative algorithm for the network is shown below. For $t = 0$, $H(0)$ consists of eight nodes forming a hexahedron. Then, we add four nodes into one face of the hexahedron. These four new nodes are linked to each other shaping a new quadrangle, a node of the new quadrilateral is connected to a node in the original quadrilateral. Thus we get $H(1)$, see Figure 3.
\begin{figure*}
	\centering
	\includegraphics[height=2.5cm,width=6.9cm]{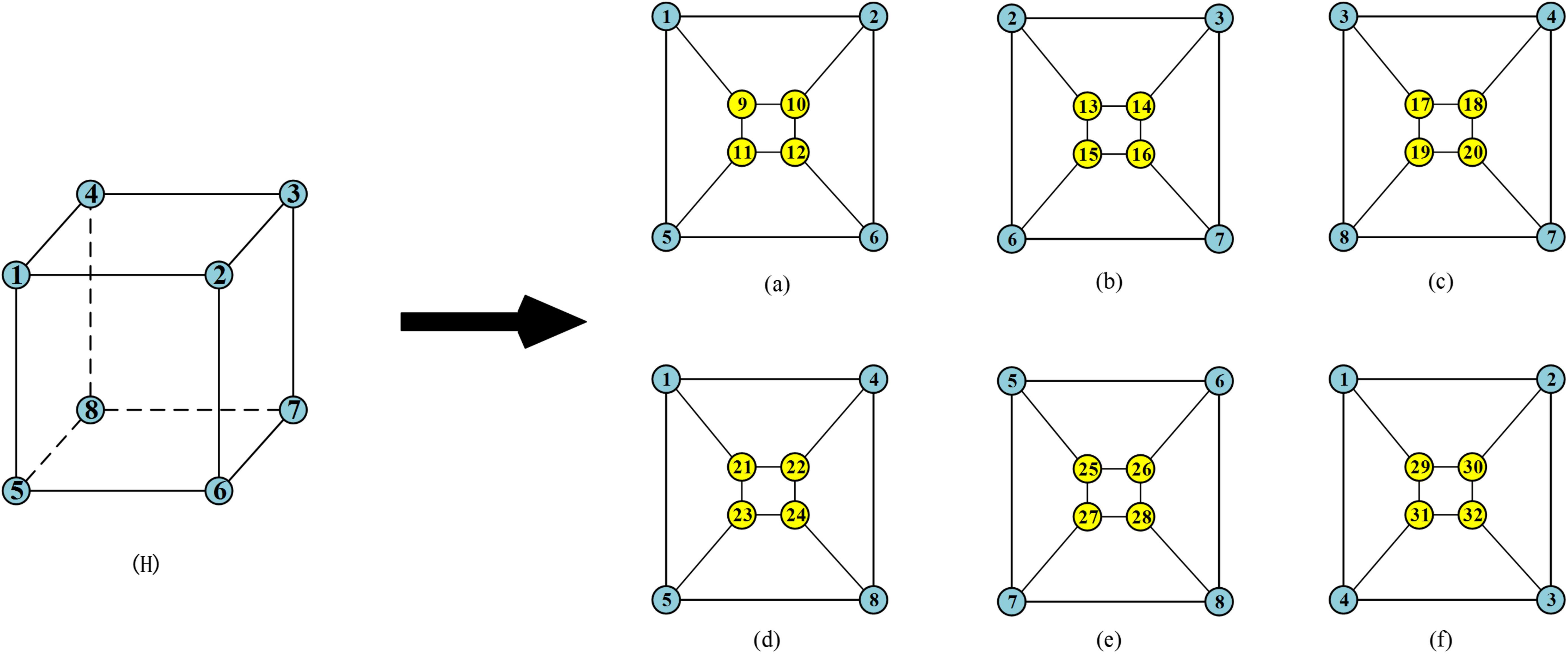}
	\caption{(Color online) Iterative construction method for the network}
	\label{num_histo_pic}
\end{figure*}

Now we compute the order and size (number of all edges) of the network $H(t)$. Let $L_{v}(t)$, $L_{e}(t)$ and $L_{\square}(t)$ be the number of vertices, edges and active triangles created at step $t$, respectively. By construction (see also Fig. 3),each active quadrangle in $H(t-1)$ will be added one quadrangle in $H(t)$. So we can find the following relationship: $L_{\square}(t) = L_{\square}(t-1)\times5$. Since $L_{\square}(0) = 6$,we have $L_{\square}(t) = 6\times5^t$.

Each active quadrilateral of $H(t-1)$ in the iteration will result in adding four new nodes and eight new edges at $t$,and then we get the following relationship:$L_{v}(t) = 4\times L_{\square}(t) = 24\times5^t$, and $L_{e}(t) = 8\times L_{\square}(t) = 48\times5^t$ for arbitrary $t > 0$. From these results, we can compute the order and size of the network. The total number of vertices $N_{t}$ and edges $E_{t}$ present at step $t$ is
\begin{equation}
N_{t}=\sum _{t_i=0}^t L_{v}(t_{i})+8=6\times5^{t+1}+2
\end{equation}

and
\begin{equation}
E_{t}=\sum _{t_i=0}^t L_{e} (t_{i})+12=12\times5^{t+1},
\end{equation}
respectively. So for large t, the average degree $\overline{k}_{t}=\frac{2E_{t}}{N_{t}}$ is approximately 4,which shows the network is sparse as most real systems.

from equations(1) and (2),we have
\begin{equation}
E_{t}=12\times5^{t+1}=2\times(6\times5^{t+1}+2)-4
\end{equation}

and
\begin{equation}
E_{t}=2N_{t}-4,
\end{equation}
On the other hand, any two edges that obtained by each iteration during network construction do not intersect each other.Thus this network is a maximal planar graph[15], which is similar to some previously studied networks [16-19].
\section{Topological properties of the network}
In this section, we will study some relevant characteristics of network $H(t)$, which are degree distribution, clustering coefficient and degree correlation.
\subsection{Degree distribution}
When a new node $i$ is added to the network at step $t_{i} (t_{i} \geq1)$, it has a degree of 3.Let $L_{\square}(i, t)$ be the number of active quadrangle at step $t$ that will create new nodes connected to the node $i$ at step $t+1$. Then at step $t_{i}$,$L_{\square}(i, t_{i})= 3$. From the iterative generation process of the network, one can see that at any subsequent step each a new neighbors of $i$ generate two active quadrangle involving $i$, and one of its existing active quadrangle is deactivated simultaneously. We define $k_{i}(t)$ as the degree of node $i$ at time $t$, then the relation between $k_{i}(t)$ and $L_{\square}(i, t)$ satisfies:

\begin{equation}
L_{\square}(i,t)=k_{i}(t)
\end{equation}

Now we compute $L_{\square}(i, t)$. By construction, $L_{\square}(i, t) = L_{\square}(i, t-1)$. Considering the initial condition $L_{\square}(i, t_{i}) =3$, we can derive $L_{\square}(i, t) =3\times2^{(t-t_{i})}$. Then at time $t$, the degree of vertex $i$ becomes
\begin{equation}
k_{i}(t)=3\times2^{(t-t_{i})}
\end{equation}
It should be mentioned that the initial eight vertices created at step 0 have a little different evolution process from other ones. We can easily obtain that at step t, the degree of one of the initial eight vertices and the number of active quadrangle are $3\times2^{t}$ .

Equation(6) shows that the degree spectrum of the network is discrete.It follows that the cumulative degree distribution[2] is given by
\begin{equation}
P_{\rm cum}(k)=\sum_{\tau \leqslant t_{i}} \frac{L_{v}(\tau)}{N_{t}}=\frac{6\times5^{t_{i}+1}+2}{6\times5^{t+1}+2}
\end{equation}
Substituting for $t_{i}$ in this expression using $t_{i}=t+\frac{\rm ln3-lnk}{\rm ln2}$ gives

\begin{equation}
P_{\rm cum}(k)=\frac{6\times5^{t+\frac{\rm ln3-lnk}{\rm ln2}}+2}{6\times5^{t}+2}
\end{equation}
When $t$ is large enough,one can obtain
\begin{equation}
P_{\rm cum}(k)=k^{(-\rm ln5/ln2)}
\end{equation}
So the degree distribution follows a power law form with the exponent $\gamma=1+\frac{\rm ln5}{\rm ln2}$.Note that the same degree exponent has been obtained in some other deterministic models such as Apollonian networks[16-17,19] and pseudofractal scale-free web[20-22].
\subsection{Clutering coefficient}
The clustering coefficient [4] of a node i with degree $k_{i}$ is given by $C_{i} = 2e_{i}/[k_{i}(k_{i}-1)]$, where $e_{i}$ is the number of existing edges among the $k_{i}$ neighbors. Using the connection rules, it is straightforward to calculate analytically the clustering coefficient $C(k)$ for a single node with degree $k$. When a node is added into the network,$k_{i}$ is 3, $e_{i}$ is zero. At each subsequent discrete time step, each of its active quadrangles increases both $k_{i}$ and $e_{i}$ by 3 and 0,respectively. therefore $C_{i}=0$,so this network is highly sparse.

\subsection{Degree correlations}
Degree correlation is a particularly interesting subject in the field of network science [23-27],because it can give rise to some interesting network structure effects.An interesting quantity related to degree correlations is the average degree of the nearest neighbors for nodes with degree $k$,denoted as $k_{nn}(k)$,which is a function of node degree $k$[24,25].When $k_{nn}(k)$ increases with $k$,it means that nodes have a tendency to connect to nodes with a similar or larger degree .In this case the network is defined as assortative [26,27].In contrast,if $k_{nn}(k)$ is decreasing with $k$, which implies that nodes of large degree are likely to have near neighbors with small degree ,then the network is said to be disassortative.If correlations are absent,$k_{nn}(k)=const$.

We can exactly calculate $k_{nn}(k)$ for the networks using equations (3) and(4) to work out how many links are made at a particular step to nodes with a particular degree.By construction,we have the following expression[28,29]
\begin{equation}
k_{\rm nn}(k)=\frac{1}{L_{v}(t_{i})k(t_{i},t)}\Bigg(\sum_{t^{'}_{i}=0}^{t_{i}^{'}=t_{i}-1}L_{v}(t_{i}^{'})L_{\square}(t_{i}^{'},t_{i}-1)k(t_{i}^{'},t)\nonumber \end{equation}

\begin{equation}
+\sum_{t_{i}^{'}=t_{i}+1}^{t_{i}^{'}=t}L_{v}(t_{i})L_{\square}(t_{i},t_{i}^{'}-1)k(t_{i}^{'},t)\Bigg)
\end{equation}

for $k=3\times2^{t-t_i}$ and where $k(t_{i},t)$ is the degree of a node $i$ at time $t$ that was born at step $t_{i}$.Here the first sum on the right-hand side accounts for the links made to nodes with larger degree(i.e. $t_{i}^{'}<t_i$)when the node was generated at $t_{i}$.The second sum describes the links made to the current smallest degree nodes at each step $t_{i}^{'}>t_{i}$.After some algebraic manipulations,we can rewrite equation(19) in term of $k$ to obtain

\begin{equation}
k_{\rm nn}(k)=\frac{690\times5^{t}-36\times2^{t}\times5^{\frac{\rm lnk-ln3}{\rm ln2}-t}-218\times5^{\frac{\rm ln3-lnk}{\rm ln2}+t}\times k}{23k}
\end{equation}

Here,$k_{nn}(k)$ increases with $k$ in equation (11),it means that network nodes have a tendency to connect to nodes with similar or larger degree. Therefore,Network is assortative.

\subsection{Conclusion}
In conclusion, based on the deterministic bubble networks(DBNs), we constructed the plane network by superimposing regular hexahedron, and analyzed the topology properties of the network.The analytic expressions of the main topological properties have been obtained through the iterative method, such as degree distribution, clustering coefficient and degree correlation.Research shows that this network shows some important features of real systems: power-law degree distribution, highly sparse and positive degree correlation.

This work opens up new interesting topics for future research and research on complex networks.In this paper, the method of iterative loop superposition of hexahedron can produce new planar graphs, which is also applicable to polyhedra and they are therefore of general applicability.

%\bibliography{myreference}

\end{document}